\documentclass[preprintnumbers,twocolumn]{revtex4-1}

\usepackage{graphicx}
\usepackage{bm}
\usepackage{amsfonts}

\def\Tr {\mathop{\hbox{Tr}}}

\newcommand{\beq}{\begin{equation}}
\newcommand{\eeq}{\end{equation}}
\newcommand{\beqa}{\begin{eqnarray}}
\newcommand{\eeqa}{\end{eqnarray}}

\newcommand\comment[1]{}

\begin{document}

\preprint{UK/11-02, NT@UW-11-25}

\title{Comment on `Controversy concerning the definition of quark and gluon angular momentum'
by Elliot Leader (PRD 83, 096012 (2011))}

\author{Huey-Wen Lin}
\email{hwlin@phys.washington.edu}
\affiliation{Department of Physics, University of Washington, Seattle, Washington 98195, USA}
\author{Keh-Fei Liu}
\email{liu@pa.uky.edu}
\affiliation{Department of Physics and Astronomy, University of Kentucky, Lexington, Kentucky 40506, USA}

\begin{abstract}
It is argued by the author that the canonical form of the quark energy-momentum tensor with
a partial derivative instead of the covariant derivative is the correct definition for the quark
momentum and angular momentum fraction of the nucleon in covariant quantization. Although it is not manifestly gauge
invariant, its matrix elements in the nucleon will be non-vanishing and are gauge invariant.
We test this idea in the path-integral quantization by calculating correlation functions on the
lattice with a gauge-invariant nucleon interpolation field and replacing the gauge link in the quark lattice
momentum operator with unity, which corresponds to the partial derivative in the continuum. We find that the
ratios of three-point to two-point functions are zero within errors for both the $u$ and $d$ quarks, contrary to
the case without setting the gauge links to unity.
\end{abstract}

\pacs{12.38.-t, 11.15.Ha, 12.38.Gc}

\keywords{}
\maketitle

     There is an ongoing controversy in QCD over the issue of how to separate total
angular momentum between the quark and glue sectors and, moreover, whether there exists a
gauge-invariant decomposition of the glue angular momentum into spin and orbital angular
momentum components as in the quark case. While several versions of gauge-invariant separation of
the glue spin and orbital angular momentum have been proposed~\cite{cls08,wak10,cgz10,wws10}, X.~Ji argued
that such a separation cannot be be achieved~\cite{ji10}. On the other hand, E. Leader asserts 
~\cite{lea11} that, if the momentum operators are to be the generators of spatial translations, then the 
gauge-invariant Bellinfante momentum operators are not correct and that rather it is the canonical form that is correct.

     Since this approach is based on covariant quantization, we shall use the path-integral quantization
to check this conclusion. In the Monte Carlo simulation of QCD on the lattice, the number of degrees of freedom is
finite, and one does not need to fix the gauge. In this case, it is usually presumed that gauge-variant observables
vanish according to Elitzur's theorem. However, E.~Leader suggests~\cite{lea11a} that, despite the
fact that the operator is not manifestly gauge invariant, there could be a gauge-invariant part of the
matrix element. To check this possibility, we perform a lattice gauge calculation of the quark energy-momentum
tensor operator. To calculate the quark momentum fraction in the nucleon on the lattice, one typically
considers the ratio of the 3-point to 2-point function
\begin{equation} \label{x}
\frac{1}{p_i} \frac{\Tr \Gamma G_{N\mathcal{O}N}(t_f,t,t_0; p_i)}{\Tr \Gamma G_{NN}(t_f,t_0; p_i)}
\stackrel{t_f - t \gg 1, t -t_0 \gg 1}{\hrulefill_{\bf \longrightarrow}} \langle x\rangle,
\end{equation}
where $\Gamma$ is the projector onto the unpolarized nucleon state. The two-point and three-point functions for the
nucleon are
\begin{eqnarray}
&{}&G_{NN}^{\alpha \beta}(t_f, t_0; p_i)= \nonumber \\
&{}&\sum_{\vec{x}} e^{-i p_i(x - x_0)} \langle \chi^{\alpha}(\vec{x},t_f)
\bar{\chi}^{\beta}(\vec{x_0}, t_0) \rangle,
\end{eqnarray}
%
%
\begin{eqnarray}
&{}&G_{N\mathcal{O}N}^{\alpha\beta}(t_f,t,t_0; p_i) =   \nonumber \\
&{}& \sum_{\vec{x}_f, \vec{x}} e^{-i p_i(x_f - x_0)}
\langle \chi^{\alpha}(\vec{x_f},t_f) \mathcal{O}(\vec{x}, t) \bar{\chi}^{\beta}(\vec{x_0}, t_0) \rangle,
\end{eqnarray}
where $\chi$ is the nucleon interpolating field and $t_0/t_f$ is the source/sink time of the nucleon
interpolation field. The lattice operator $\mathcal{O}$ for the momentum that has been calculated in the
literature~\cite{dsd09} is
\begin{equation}  \label{t4i}
T_{4i}(x) = \frac{-i}{8a}\frac{1}{2\kappa}[\bar{\psi}(x)\gamma_4 U_i(x)\psi + ...]
\end{equation}
\begin{figure}[h!]
  \center
\includegraphics[width=3.0in]{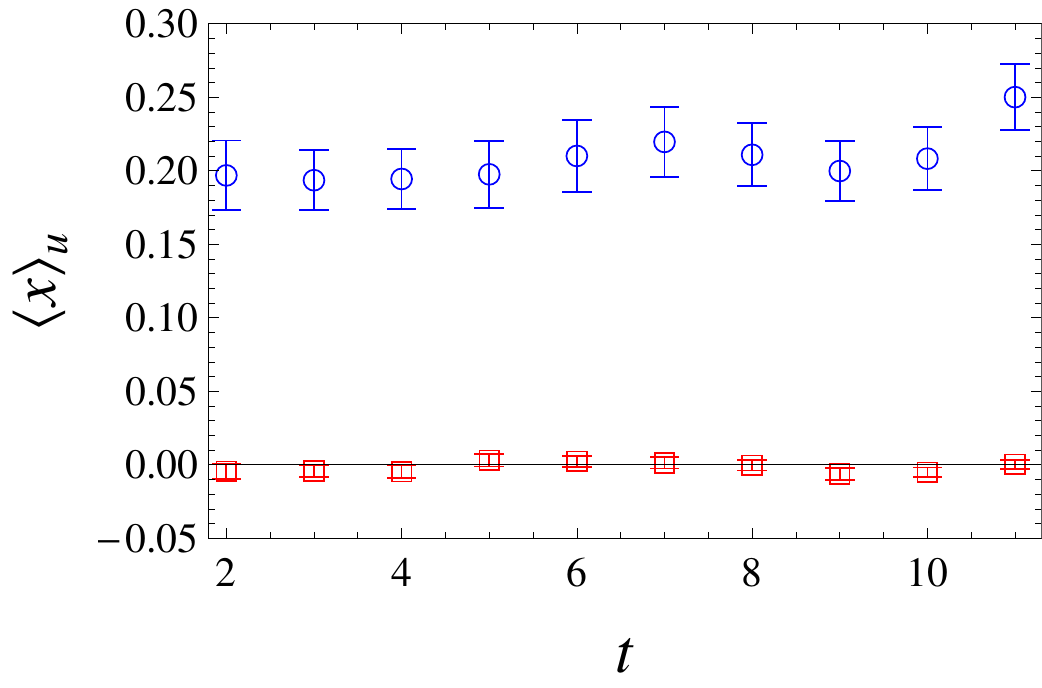}
\includegraphics[width=3.0in]{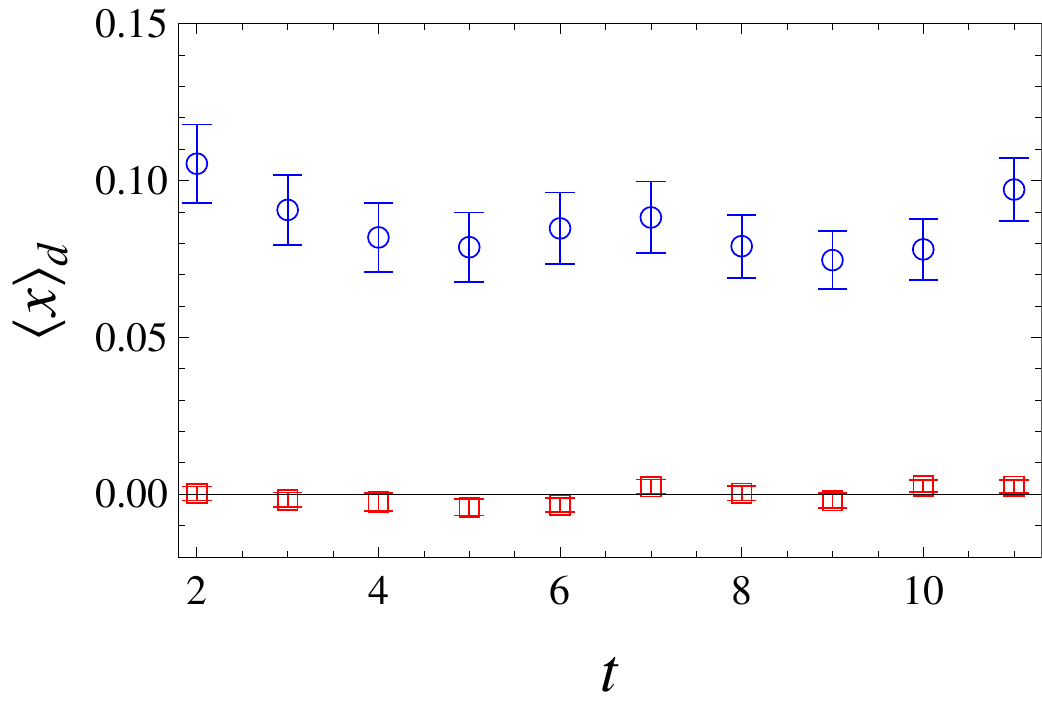}
\caption{The ratio of the three-point to two-point functions in Eq.~(\ref{x}) is plotted as a function of the
operator insertion time $t$ for the $u/d$ quark in the upper/lower panel. The blue circles are results from the
momentum operator with the covariant derivative and the red squares are those corresponding to the partial
derivative.}\label{fig:x}
\end{figure}
with $U_i(x)$ being the gauge link. This corresponds to the continuum tensor operator for the quark momentum
\mbox{$P_i = \frac{-i}{4}\bar{\psi}(\gamma_4\!\! \stackrel{\leftrightarrow}{D}_i + \gamma_i\!\!
\stackrel{\leftrightarrow}{D}_4)\psi$} which is of the Bellinfante form. E.~Leader argues~\cite{lea11} that the correct
momentum operator should be
the canonical form, with the covariant derivative in the Bellinfante form replaced with the partial derivative.
This corresponds to setting the gauge link $U$ to unity in the lattice operator $T_{4i}(x)$ in Eq.~(\ref{t4i}).

We performed a lattice calculation on a $2+1+1$-flavor $24^3 \times 64$ HISQ lattice~\cite{Bazavov:2010ru} (HYP-smeared) 
with the Wilson-clover valence fermions.
The pion mass is 310 MeV with lattice spacing around 0.12~fm, and 127 configurations (each with 4 sources) were used 
in the calculation. The nucleon
source is shifted to $t_0 = 0$ and sink timeslice is at $t_f = 12$. The ratio in Eq.~(\ref{x}) for the connected insertion
is plotted in Fig.~\ref{fig:x} as a function of the operator insertion time $t$ for the $u$ quark in the upper panel and
the $d$ quark in the lower panel. The results with the Bellinfante form are shown as blue circles; they are
obviously non-zero. The quark momentum fraction $\langle x\rangle$ is usually extracted from a plateau in a window
away from the source and the sink. On the other hand, when the gauge link $U$ is set to unity, which corresponds to
a partial derivative, the results (red squares) are zero within errors for practically all insertion times
$t$. We conclude from this calculation that the canonical form of the quark energy-momentum tensor with partial
derivatives does not lead to non-zero matrix elements in the nucleon.

This work is partially supported by U.S. DOE Grant No. DE-FG05-84ER40154. HWL is supported by the DOE grant DE-FG02-97ER4014. 
The numerical work is performed on Hyak clusters at the University of Washington eScience Institute, using hardware 
awarded by NSF grant PHY-09227700.

\end{document}